\documentclass[aps,prb,reprint,superscriptaddress,longbibliography,nofootinbib,noeprint]{revtex4-2}

\usepackage[T1]{fontenc}
\usepackage{mathtools,amssymb,slashed,bbm,tensor,verbatim,siunitx,slashed}
\usepackage{tikz}
\usetikzlibrary{decorations.markings,decorations.pathreplacing,arrows,arrows.meta,calc,shapes.arrows,shapes.geometric}
\usepackage{array,multirow}
\usepackage[colorlinks, linkcolor=red,citecolor=blue,urlcolor=blue]{hyperref}
\usepackage[all]{hypcap}

\usepackage[caption=false]{subfig}

\DeclareRobustCommand{\a}{a_{\textsc{b}}}

\usepackage{xprintlen}

\begin{document} 
\title{Large inverse Faraday effect for Rydberg states of free atoms and isolated donors in semiconductors}

\author{Patrick J. Wong}
\affiliation{Nordita, Stockholm University and KTH Royal Institute of Technology, Hannes Alfv\'ens v\"ag 12, SE-106~91 Stockholm, Sweden}
\affiliation{Department of Physics, University of Connecticut, Storrs, Connecticut 06269, USA}

\author{Ivan M. Khaymovich}
\affiliation{Nordita, Stockholm University and KTH Royal Institute of Technology, Hannes Alfv\'ens v\"ag 12, SE-106~91 Stockholm, Sweden}

\author{Gabriel Aeppli}
\affiliation{Laboratory for Solid State Physics and Quantum Center, ETH Z\"urich, CH-8093 Z\"urich , Switzerland}
\affiliation{Institut de Physique, EPFL, CH-1015 Lausanne, Switzerland}
\affiliation{Photon Science Division, Paul Scherrer Institut, CH-5232 Villigen, Switzerland}

\author{Alexander V. Balatsky}
\affiliation{Nordita, Stockholm University and KTH Royal Institute of Technology, Hannes Alfv\'ens v\"ag 12, SE-106~91 Stockholm, Sweden}
\affiliation{Department of Physics, University of Connecticut, Storrs, Connecticut 06269, USA}

\date{\today}

\begin{abstract}
We report on the induction of magnetization in Rydberg systems by means of the inverse Faraday effect, and propose the appearance of the effect in two such systems, Rydberg atoms proper and shallow dopants in semiconductors. 
Rydberg atoms are characterized by a large orbital radius. This large radius gives such excited states a large angular moment, which when driven with circularly polarized light,  translates to a large effective magnetic field $\mathcal{B}_{\text{eff}}$. We calculate this effect to generate effective magnetic fields of $O(\qty{1}{\micro\tesla})\times\left( \frac{\omega}{\qty{1}{\tera\hertz}} \right)^{-1} \left( \frac{\mathcal{I}}{\qty{10}{\watt\per\centi\meter\squared}} \right) n^4$ in the Rydberg states of atoms such as Rb and Cs for off-resonant photon beams with frequency omega and intensity $\mathcal{I}$ expressed in units of the denominators and $n$ the principal quantum number. 
Additionally, terahertz spectroscopy of phosphorus doped silicon reveals a large cross-section for excitation of shallow dopants to Rydberg-like states, which even for small $n$ have the potential to be driven similarly with circularly polarized light to produce an even larger magnetization. Our theoretical calculations estimate $\mathcal{B}_{\text{eff}}$ as $O(\qty{E2}{\tesla})$ for Si:P with a beam intensity of $\SI{E8}{\watt\per\centi\meter\squared}$.
\end{abstract}

\maketitle

\section{Introduction}

The inverse Faraday effect (IFE) is a well-known optomagnetic phenomenon in which a static (dc) magnetization is dynamically induced in matter by a light field~\cite{pitaevskii,landaulifshitz,pershan1963,vanderziel,pershan,kirilyuk,battiato}. It is often stated as an induction of dc magnetization as a result of illumination by circularly polarized light:
\begin{equation}
    \vec{M}_{\text{dc}} \sim \vec{E}(\omega) \times \vec{E}^*(\omega) \,.
\label{Eq:1}
\end{equation} Here  $\vec{E}(\omega)$ is the complex-valued electric field of a light vector at frequency $\omega$, which we assume to be monochromatic hereafter. 

The inverse Faraday effect (IFE) is in contrast to the conventional Faraday effect. In the latter, a linearly polarized light passing through a magnetized medium undergoes a rotation in its polarization axis. On the other hand, the IFE takes place when circularly polarized light, with a rotating polarization, induces a magnetization, denoted by $\vec{M}_{\text{dc}}$. One particularly intriguing application of the IFE in magnetic systems is ultrafast magnetic switching~\cite{lottermoser,kimel}. The IFE has been observed and studied in a plethora of systems, from various materials and molecules to nanostructures such as gold nanoparticles~\cite{cheng2020}.
The IFE has been proposed to be seen in metals and superconductors \cite{putilov,mironov,croitoru,plastovets,majedi,sharma}, Mott insulators~\cite{banerjee}, non-magnetic compounds~\cite{gukornev}, and Dirac and Weyl semimetals~\cite{taguchi,zyuzin,tokman,liang}. 

Here we explore the IFE in Rydberg systems, including both free atoms and shallow semiconductor dopants. One notable characteristic of the IFE is its ability to induce significant orbital magnetization within the medium.
This has potential applications in the realm of free-atom and dopant-based quantum computing~\cite{greenland,fricke} and manipulation of the metal-insulator transition in doped semiconductors~\cite{sarachik}. In these cases the IFE can be utilized to coherently manipulate the magnetic states of the system.

\begin{figure}[h]
\centering
\includegraphics[scale=1]{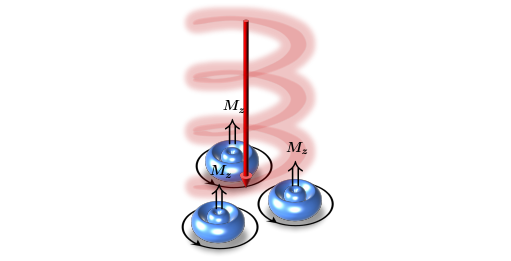}
\caption{Schematic of induced orbital angular momentum generating magnetic moment. Illustrated are excited atomic wave functions, characteristic of either the dopant states in a semiconductor or Rydberg atoms. The incoming circularly polarized beam induces an emergent ferromagnetic state with oriented magnetic moments $M_z$ arising from the driven orbital angular momentum. 
\label{fig:schematic}}
\end{figure}

At its core, the strength of the effect is linked to the size of an electron's bound state. In simpler terms, the induced magnetic moment can be understood as the induced angular momentum of an orbital particle's movement.
\begin{equation}
    \vec{M}_{\text{dc}} \sim \langle \vec{r} \times \partial_t \vec{r} \rangle .
\end{equation}{}  
Further, the effect scales as the squared effective Bohr radius $\a$:
\begin{equation}
    M_{\text{dc}} \sim \a^2 \,.
\end{equation} 
Here $\vec{r}$ is the operator of the electron center of mass and we estimate the average of the electron angular momentum scale as $\a$. Thus, the larger the effective radius of the bound state is, the greater the induced moment becomes. This insight led us to hypothesize a great amplification of the IFE as we go to higher-lying Rydberg states in free atoms. The effect is four orders of magnitude larger for shallow dopants in semiconductors, where the Bohr radius, which when multiplied with $n^2$ sets the length scale for all Rydberg states with principal quantum number $n$, exceeds, by roughly two orders of magnitude, that for the conventional hydrogen atom (Fig.~\ref{fig:schematic}).

Study of the IFE in Rydberg states is lacking. The IFE for the hydrogen atom has been briefly discussed in Ref.~\cite{battiato}; however, it is not applicable to Rydberg states as their analysis only considers the lowest principal quantum numbers and stops already at $n=3$.

The IFE is characterized by a nonlinear optical response which can be described by $M^a(0) = \chi^{abc}(0;\omega,-\omega) E_b(\omega) E_c(-\omega)$, where the indices $a$, $b$, and $c$ denote Cartesian coordinates and the Einstein summation convention is implied.
Previous work~\cite{sipe} has calculated the general second-order optical response $\chi^{abc}(\omega;\omega',\omega'')$ in bulk semiconductors, including the rectified zeroth harmonic generation susceptibility $\chi^{abc}(0;\omega,-\omega)$. 
In bulk semiconducting systems this response tensor also contributes to the photovoltaic effect \cite{vonbaltz} and shift photocurrent \cite{fregoso}. 
The second-order susceptibility tensor obeys the symmetry condition $\chi^{abc} = - \chi^{acb}$, meaning that it is only non-zero for systems without an inversion symmetry in a static case. Dynamically, circularly polarized light itself lowers the symmetry of the system so that to second order we generate dc magnetization even if the inversion is not broken in equilibrium. A good discussion of the second-order effects and symmetry requirements is given in Ref.~\cite{sipe}.
The nonlinear response theory studied in these previous works focused on the response of bulk materials. The focus of the present paper is not on the properties of the bulk semiconductor, but rather on the magnetic state of free atoms in vacuum and atomic like dopants in semiconductors.


The structure of the paper is as follows. We review the heuristic classical picture and quantum theory of the inverse Faraday effect in Sec.~\ref{sec:ife}. In Sec.~\ref{sec:rydberg} we discuss the IFE in Rydberg atoms. In Sec.~\ref{sec:shallowdopants}  we present the analysis of the IFE for shallow dopants in semiconductors. In Sec.~\ref{sec:conclusion} we provide a summary. Expanded technical details and additional numerical data are given in the Appendixes.

\section{Inverse Faraday effect\label{sec:ife}}

 We first expand upon the heuristic classical model of the effect given in the Introduction before reviewing the quantum description developed in Ref.~\cite{pershan}. Expanded details of the development of the quantum theory are in Appendix~\ref{sec:derivation}.

\subsection{Classical model}
Classically, the inverse Faraday effect can be understood as the magnetic moment arising from the orbital angular momentum of a charged particle per unit mass as
\begin{equation}
	\vec{M} = \frac{q}{2 m_*} \vec{r}(t) \times \vec{p}(t) = \frac{q}{2} \vec{r}(t) \times \partial_t \vec{r}(t) ,
\label{eq:basic}
\end{equation}
where $q$ is the electric charge of this particle and $m_*$ its effective mass.
This expression illustrates that the magnitude of the induced magnetic moment is proportional to the square of the orbital radius $|\vec{r}(t)|^2$. In the context of Rydberg atoms and shallow semiconductor dopants discussed above, the large effective Bohr radius they possess therefore implies the possibility of inducing a large magnetization due to the inverse Faraday effect.

The magnetic moment in Eq.~\eqref{eq:basic} can be obtained by considering free electrons under the Drude-Lorentz model~\cite{battiato}. In this model, the equation of motion for the electrons is given by
\begin{equation}
    \partial_t^2 \vec{r}(t) + \gamma \partial_t \vec{r}(t) + \omega_0^2 \vec{r}(t) = \frac{\vec{F}(t)}{m_*} \,.
\end{equation}
For circularly polarized light of frequency $\omega$, the external force takes the form of
\begin{equation}
\begin{multlined}
    \vec{F}(t) = q \vec{E}(t) = q \mathcal{E} \left[ \cos(\omega t) \hat{x} + \sin(\omega t) \hat{y} \right] \\\equiv q\, \Re\left[ \vec{E}(\omega) e^{-i \omega t} \right] \,,
\end{multlined}
\end{equation}
with the amplitude $\mathcal{E}$ and complex-valued vector $\vec{E}(\omega) = \mathcal{E} \left[ \hat{x} + i \hat{y} \right]$ of the external monochromatic electric field, and unit vectors $\hat{x}$ and $\hat{y}$ in Cartesian coordinates.
The solution in the frequency domain $\vec{r}(t) = \Re\left[ \vec{r}(\omega) e^{-i \omega t} \right]$ for the equation of motion reads as
\begin{equation}
    \vec{r}(\omega)
    =
    \frac{q}{m_*} \frac{1}{(\omega^2 - \omega_0^2) + i\gamma \omega} \vec{E}(\omega) \,.
\end{equation}
This results in a magnetic moment induced by the IFE as given by Eq.~\eqref{eq:basic} as
\begin{equation}
\begin{aligned}[b]
    \vec{M} 
    &=  \frac{q}{2}\, \vec{r}(t) \times \partial_t \vec{r}(t)
    \\
    &=  \frac{q}{8} \left[ \vec{r}(\omega) e^{-i\omega t} + \text{c.c.} \right] \times \left[ i \omega \vec{r}^*(\omega) e^{i\omega t} + \text{c.c.} \right]
    \\
    &=  \frac{q}{4}\, \Im \left[ \vec{r}(\omega) \times \omega \vec{r}^*(\omega) \right]
    \\
    &= \frac{q^3}{4 m_*^2} \frac{\omega}{(\omega^2 - \omega_0^2)^2 + \gamma^2 \omega^2} \Im \left[ \vec{E}(\omega) \times \vec{E}^*(\omega) \right]
\end{aligned}
\end{equation}
To model a bound state, we consider a coherent light source with frequency at the resonance $\omega_0$, when the damping term dominates in the denominator. The induced magnetic moment then takes the form
\begin{equation}
    {M}_z \approx \frac{q^3}{2 m_*^2} \frac{\mathcal{E}^2}{\gamma^2 \omega} \,.
\end{equation}
With a damping dominated term, $1/\gamma$ is proportional to the average radius $\langle r \rangle$ as $r(\omega) \sim 1/(\omega^2 + i \gamma \omega - \omega_0^2)$ in an undriven system. We identify this with the Bohr radius $\a$. We then interpret the magnetic moment as
\begin{equation}
    M_z \propto \frac{\a^2 \mathcal{E}^2}{\omega} \,.
\end{equation}
Here we see the functional dependence of the induced magnetization on $\a$, $\omega$, and $\mathcal{E}$. This scaling dependence will also be seen in the quantum calculation.

\subsection{Quantum theory \label{sec:quantumife}}

The quantum mechanical description of the IFE was developed in Ref.~\cite{pershan} as an effect arising from a second-order process in the dipole interaction.
The Hamiltonian for this system is given by $H = H_0 + V(t)$ with the potential
\begin{equation}
	V(t) = v(t) e^{i \omega t} + v^*(t) e^{-i \omega t}
\end{equation}
where $v(t) = - q\, \vec{r} \cdot \vec{E}(t)$. The applied field $\vec{E}$ is taken to be circularly polarized and propagating in the $\hat{z}$ direction, which is defined to be the direction perpendicular to the target sample surface.  
Defining $r_\pm = \frac{1}{\sqrt{2}} ( x \pm i y)$ and $\mathcal{E}_{L/R} = \frac{1}{\sqrt{2}}( \mathcal{E}_x \mp i \mathcal{E}_y )$, we have $v = -q \left( r_+ \mathcal{E}_R + r_- \mathcal{E}_L \right)$, where $L$ and $R$ denote left- and right-hand circular polarizations.
Following Ref.~\cite{pershan}, an effective Hamiltonian can be defined to describe the second-order dipole interaction process yielding the IFE.
The effective Hamiltonian can be expressed as an operator whose matrix elements are given by
\begin{equation}
    \langle a | \overline{H}_{\text{eff}}(t) | b \rangle
    \vcentcolon=  \frac{i}{\hbar} \sum_{c} \langle a | \overline{V}(t) | c \rangle \int_{-\infty}^{t} d t' \langle c | \overline{V}(t') | b \rangle
\end{equation}
where $\overline{f}(t) = e^{i H_0 t/\hbar} f(t) e^{-i H_0 t/\hbar}$, $|a\rangle$ and $|b\rangle$ are eigenstates of the ground multiplet with energies $E_a$ and $E_b$, and the summation is taken over the excited states $|c\rangle$ with energies $E_c$.
These matrix elements for the zeroth harmonic generation components for the dipole interaction can be computed as
\begin{equation}
\begin{multlined}[c]
	\langle a | H_{\text{eff}} | b \rangle
	= \frac{q^2}{\hbar} \biggl[ \omega \left( \mathcal{E}_R^2 - \mathcal{E}_L^2 \right)
    \times\\\times
	\sum_c \left( \frac{\langle a |r_+| c \rangle \langle c |r_-| b \rangle }{\omega_{bc}^2 - \omega^2} - \frac{\langle a |r_-| c \rangle \langle c |r_+| b \rangle }{\omega_{bc}^2 - \omega^2} \right) \biggr]
\label{eq:effham}
\end{multlined}
\end{equation}
where $\hbar \omega_{bc} = E_b - E_c$, and $r_\pm$ and $\mathcal{E}_{L/R}$ defined as above.
Details on the derivation of the expression Eq.~\eqref{eq:effham} can be found in Refs.~\cite{pershan,popova} as well as Appendix~\ref{sec:derivation}.
\footnote{This calculation ignores temperature dependence and a full finite-temperature calculation is beyond the scope of this work, however our effective Hamiltonian in Matsubara frequency can be schematically described as an intensity with scattering rate $\Gamma$ as $\mathcal{I}(\omega_n) = \frac{1}{(i\omega_n - \omega)^2 - \Gamma^2}$. At finite temperature this becomes $\mathcal{I}(T,\omega_n) = \mathcal{I}(0,\omega_n) / \left[1 + T^2/\Gamma^2 \right]$, so that the effect of temperature is a suppression by a factor of $\frac{1}{1 + T^2/\Gamma^2}$.}
The analysis of the effect in the present discussion is valid for long times, where the characteristic time scale of fluctuations in the applied light source $v(t)$ is on the order of nanoseconds or picoseconds. For the ultrafast regime of femtoseconds, the analysis needs to be augmented by considering the shape of the incident pulse. Details regarding modifications to the effective Hamiltonian necessary for this regime can be found in Ref.~\cite{popova}.

\begin{figure}[htp!]
    \includegraphics[scale=1]{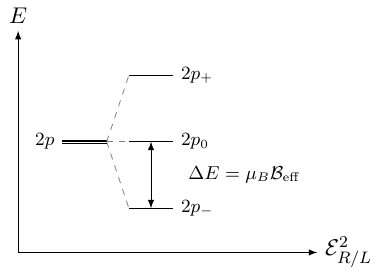}
    \caption{Splitting of degenerate energy levels due to a second-order interaction with the applied circularly polarized field $\mathcal{E}_{R/L}$. The energy splitting $\Delta E$ is attributed to the effective magnetic field $\mathcal{B}_{\text{eff}}$.\label{fig:energysplitting}}
\end{figure}

The action of the effective Hamiltonian is to break the degeneracy in the energy levels between the initial and final states as illustrated in Fig.~\ref{fig:energysplitting}.
Given that the external field $\mathcal{E}_R^2 - \mathcal{E}_L^2$ in Eq.~\eqref{eq:effham} breaks time-reversal symmetry, the effective Hamiltonian can be interpreted as an effective magnetic field with an effective magnetic moment. The matrix element \eqref{eq:effham} can therefore be reexpressed in terms of this effective magnetic field instead of the physical electric field, reading
\begin{equation}
\begin{multlined}[c]
	\langle a | H_{\text{eff}} | b \rangle
	= \mu_B \biggl[\frac{2 m_* q}{\hbar^2} \omega \left( \mathcal{E}_R^2 - \mathcal{E}_L^2 \right) \times\\ \times 
	\sum_c \left( \frac{\langle a |r_+| c \rangle \langle c |r_-| b \rangle }{\omega_{bc}^2 - \omega^2} - \frac{\langle a |r_-| c \rangle \langle c |r_+| b \rangle }{\omega_{bc}^2 - \omega^2} \right)\biggr] 
\label{eq:effhamB}
\end{multlined}
\end{equation}
where $\mu_B = q \hbar / 2 m_*$ is the effective Bohr magneton and the quantity in the large square brackets has units of magnetic field strength. 
The effective Hamiltonian now takes the form of a Zeeman term $\langle a | H_{\text{eff}} | b \rangle = \mu_B \mathcal{B}_{\text{eff}}$.
We emphasize here that there is no real magnetic field present; however, the atoms respond as if they possess a unit Bohr magneton magnetic moment immersed in a magnetic field of strength $\mathcal{B}_{\text{eff}}$. We note, however, that the electron spins are subject to an effective magnetic field that is not identical to  $\mathcal{B}_{\text{eff}}$ but is dictated by the spin-orbit interaction for the Rydberg state.





We now apply this formalism to calculate the magnitude of the IFE for Rydberg atoms and shallow dopants in semiconductors. Since we are interested in Rydberg-like states, the states $|a\rangle$, $|b\rangle$, and $|c\rangle$ are taken to be hydrogenic states defined in terms of their principal quantum number $n$, orbital angular momentum number $\ell$, and magnetic quantum number $m$, such that the states can be labeled by the tuple $|a\rangle = |n,\ell,m\rangle$. Their specific form is given by Eq.~\eqref{eq:hydrogenicwf}. The magnetization we consider is an orbital angular momentum effect and we do not consider spin states in our analysis, which we leave for future work.

\section{Rydberg Atoms\label{sec:rydberg}}

Rydberg atoms are isolated atoms with a valence electron occupying a state with $n\gg1$, which results in a highly extended radial wave function and a large electric dipole moment~\cite{rydbergreview,shaffer}.
The large radial extent of the Rydberg wave function means that the outer electron experiences a very weak (shallow) potential. The shallowness of this potential allows Rydberg atoms to be highly sensitive to external electric fields which makes them an attractive platform for metrology ~\cite{adams}, with recent work demonstrating sensitivity to electric fields on the order of $\qty{E-1}{\volt\per\meter}$~\cite{facon}. Rydberg atoms have also been shown to serve as precision radio frequency sensors, which are highly tunable to specific frequencies and are robust to noise~\cite{fancher}. Finally, on account of the strong dipole blockade associated with their large spatial extent, Rydberg atoms are useful building blocks for quantum processors~\cite{lukin2001,rydbergqi}. As we will show, in addition to electric fields, these states are in principle highly sensitive to magnetic fields as well.

We are not aware of prior analysis of IFE for Rydberg states. We thus give a microscopic description of the IFE effect for Rydberg states. The dependence of the IFE on Rydberg quantum numbers can be estimated from quasiclassical considerations. A particle of charge $q$ in a bound state of a central potential with coupling constant $k$ obeys
$p^2 / 2 m r = k q^2 / r^2$. A dependence on the principal quantum number can be obtained from the Bohr-Sommerfeld quantization condition $r \cdot p = n \hbar$, or $p = n \hbar /r$. Inserting this condition into the energy balance expression yields $r = n^2 \hbar^2 / 2 m k q^2$. The radius of the bound particle therefore scales as $n^2$. We therefore expect Rydberg states to yield a strong IFE as a consequence of their large-$n$ values.

\subsection{IFE of Rydberg atoms\label{sec:rydbergife}}

Here we calculate the magnitude of the effective magnetic field induced by the IFE for the typical Rydberg atoms rubidium ($Z=37$, $\a=\qty{235}{\pico\meter}$) and cesium ($Z=55$, $\a=\qty{260}{\pico\meter}$).
In these calculations we ignore spin-orbit coupling. Since the Rydberg states have wave functions with large spatial extent, the coupling between the valence electron and the atomic nucleus is very weak, making the omission of spin-orbit coupling a reasonable assumption.
This is in contrast to typical realizations of the IFE in solid-state magnetic systems where spin-orbit coupling is a dominant contribution to the effect.

To estimate the size of the effect, we use the expression \eqref{eq:effhamB} over Rydberg wave functions with parameters of the applied beam energy of $\mathcal{I} = \qty{10}{\watt\per\centi\meter\squared}$, corresponding to $|\mathcal{E}| = \qty{E4}{\volt\per\meter}$, and $\omega = \qty{1}{\tera\hertz}$. We calculate the value of the IFE for states with the quantum numbers $|n,1,\pm1\rangle$. The calculated values for the effective induced magnetic field for rubidium and cesium over a range of principal quantum numbers $n$ is plotted in Fig.~\ref{fig:rydbergife}. Numerical values are given in Tables~\ref{tab:ruife} and~\ref{tab:csife} in Appendix~\ref{sec:rydbergdata}.
We find that the magnitude of the IFE in Rydberg systems scales as $n^4$. This scaling behavior is expected since the spatial extent of a Rydberg wave function scales as $n^2$ and the IFE scales as the radius squared. The size of the IFE for both rubidium and cesium at $n\sim30$ is $\mathcal{B}_{\text{eff}} \sim \qty{10}{\milli\tesla}$.
For the numerical values in Fig.~\ref{fig:rydbergife} and Tables~\ref{tab:ruife} and~\ref{tab:csife}, the IFE scales as
\begin{equation}
    \frac{1}{\mu_B} \langle a | H_{\text{eff}} | b \rangle = \mathcal{B}_{\text{eff}} = \SI{1}{\micro\tesla}  \left( \frac{\omega}{\qty{1}{\tera\hertz}} \right)^{-1} \left( \frac{\mathcal{I}}{\SI{10}{\watt\per\centi\meter\squared}} \right) n^4\,. 
\end{equation}

\begin{widetext}

\begin{figure}[htp!]
    \includegraphics[scale=1]{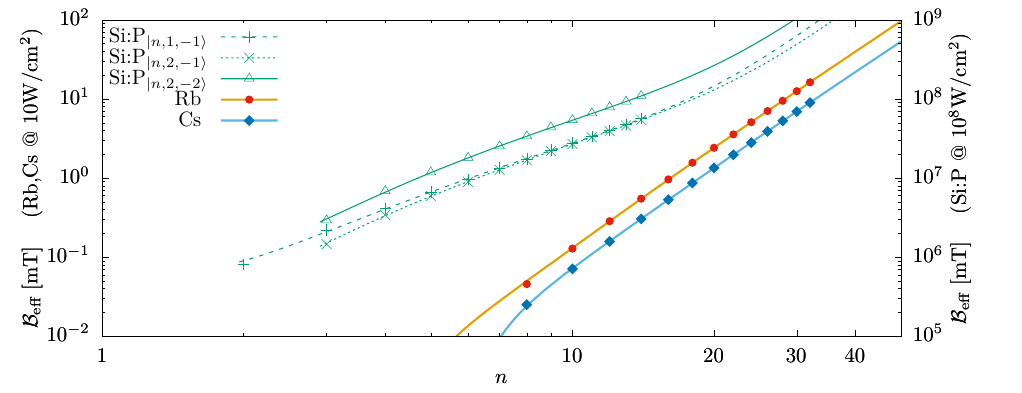}
    \caption{Magnitude of the induced effective magnetic field $\mathcal{B}_{\text{eff}}$ due to the inverse Faraday effect for Si:P (green pluses, crosses, and triangles), rubidium (red circles $\bullet$), and cesium (blue diamonds {\scriptsize$\blacklozenge$}) evaluated for a range of principal quantum numbers $n$ with an applied $\qty{1}{\tera\hertz}$ beam with an intensity of $\SI{10}{\watt/\centi\meter\squared}$ for the Rb and Cs states (left vertical axis), and a beam intensity of $\SI{E8}{\watt/\centi\meter\squared}$ for the Si:P states (right vertical axis). The data points for Si:P are for $|\ell,m\rangle = |1,-1\rangle$ ($+$), $|2,-1\rangle$ ($\times$), and $|2,-2\rangle$ ($\triangle$). The curves are fits to the numerical data and are fourth-order polynomials, given in Eqs.~\eqref{eq:sip11}--\eqref{eq:sip22} for Si:P and Eqs.~\eqref{eq:rbfit} and~\eqref{eq:csfit} for the Rydberg atoms. The curves for Si:P terminate at $n=2$ and 3 as the IFE is only finite for states where $n\geq\ell+1$. The numerical values for $\mathcal{B}_{\text{eff}}$ are shown in Table~\ref{tab:allmatrixelements} in Appendix~\ref{sec:dopantdata}, and Tables~\ref{tab:ruife} and~\ref{tab:csife} in Appendix~\ref{sec:rydbergdata}.\label{fig:rydbergife}}
\end{figure}

\end{widetext}

The above calculation shows that for Rydberg atoms, the IFE obeys $\mathcal{B}_{\text{eff}} Z^{-3/2} \approx \text{const}$. This scaling can be used to extrapolate estimates for other species of Rydberg atoms.

The IFE for the hydrogen atom was previously reported in Ref.~\cite{battiato}, where they report that the effect scales linearly with the average radial extent $\langle r \rangle$ of the wave function, or as $n^2$, in contrast with our results above. However, Ref.~\cite{battiato} only calculated the effect for up to $n=3$. For $n\leq4$, we can confirm that $n^2$ does accurately describe the scaling of the hydrogenic IFE; however, for larger values of $n$ this scaling no longer holds and $n^4$ is the appropriate scaling behavior, as we report here.

\section{Shallow Dopants\label{sec:shallowdopants}}

A Rydberg system is generally an isolated atom excited to a state with a very high principal quantum number $n\gg1$. However, due to the shallow potential of certain dopants in semiconductors, a Rydberg state can be achieved already at low $n$, such as $n=2$. This means that only a low transition energy is required to produce a widely spatially extended electronic wave function.

Recent results involving terahertz pumping of doped silicon reveal that there exists a large cross section for two photon absorption of shallow dopants to an excited Rydberg-like state~\cite{vanloon}. These Rydberg states possess a very large effective Bohr radius. The shallow potential contributing to this large radius arises from the high dielectric constant of silicon and the low effective electron mass. The effective mass of electrons in silicon is $m_* = \qty{0.3}{\mathnormal{m_e}}$ of the electron mass and the Bohr radius as observed in Ref.~\cite{vanloon,murdin2013} is $\a = 4\pi \epsilon \hbar^2 / q^2 m_* = \qty{3.17}{\nano\meter}$, which is significantly larger than a typical atomic Bohr radius such as that for hydrogen, where $\a = \qty{0.05}{\nano\meter}$, or even the Rydberg atoms considered in the preceding section which had $\a = \qty{0.235}{\nano\meter}$ and $\qty{0.260}{\nano\meter}$ for rubidium and cesium, respectively. The shallow potential of these dopant atoms means that their low-energy excited states possess a very large spatial extent with a large dipole moment.
Our goal in the present paper is to exploit this property to induce a large orbital magnetization via the IFE.

For concreteness, in this paper we consider the case of phosphorus doped silicon, Si:P. The phosphorus atoms ($Z=15$, $\a=\qty{3.17}{\nano\meter}$) form the shallow Rydberg-like states. While we focus on silicon in our present paper, we note however that compared to silicon, which has a dielectric constant of $\epsilon_{\text{Si}} = 11.8\, \epsilon_0$, germanium has a higher dielectric constant of $\epsilon_{\text{Ge}} = 16.0\, \epsilon_0$, which suggests that dopants in germanium may have an even more pronounced IFE than those for silicon as the larger dielectric constant leads to a larger Bohr radius. 


We emphasize that the electron in the excited Rydberg state is still bound to the dopant atom.  The donor electron is not excited into the continuum conduction band and remains in a bound state with the dopant atom. In the context of the present work, the doped semiconductor serves as a solid-state platform for demonstrating the IFE for individual atomic systems in contrast to the free or trapped atoms of the preceding section.


\subsection{IFE of shallow dopants\label{sec:ifeshallowdopants}}

With the expression \eqref{eq:effhamB}, we seek to evaluate the magnitude of the IFE for shallow dopants in semiconductors.
The work of Ref.~\cite{vanloon} identified a series of excited Rydberg states which are accessible from terahertz pumping of phosphorus donors in silicon. We take these states to form the basis states for the matrix elements in Eq.~\eqref{eq:effham}. 
To obtain numerical estimates of the magnitude of the effective magnetic field, we assume typical experimental values of $\qty{1}{\tera\hertz}$ for the frequency of the applied beam and a beam intensity of $\qty{E8}{\watt\per\centi\meter\squared}$, or $\qty{E7}{\volt\per\meter}$ in terms of electric field strength. This electric field strength is typical for current state-of-the-art terahertz spectroscopy experiments. 
Silicon is transparent to terahertz light, so we do not need to consider any dissipative effects. This transparency also implies that not only dopants on the surface of the silicon sample will respond to the terahertz light, but dopants within the volume will as well.
The diagonal matrix elements of the effective Hamiltonian for a selection of basis states $|a\rangle = |n,\ell,m\rangle$ yielding values for the effective magnetic field are given in Table~\ref{tab:matrixelements}. We find that for matrix elements with finite amplitude, the effective magnetic field has a magnitude of $O(\qty{E2}{\tesla})$. The total energy splitting induced by the effective Hamiltonian for the relevant states is $O(\qty{10}{\milli\electronvolt})$.
We note that the energy level splittings due to the applied field may exceed the ionization energy of the dopants. These ionization transitions are not captured by our calculation above.
Off-diagonal matrix elements are those with initial and final states differing in the principal quantum number $n$, but with same angular state $\ell,m$.
In terms of relevant parameters we see that the magnitude of the effective magnetic field scales as
\begin{equation}
    \begin{multlined}
    \frac{1}{\mu_B} \langle a | H_{\text{eff}} | b \rangle = \mathcal{B}_{\text{eff}} \\= \SI{E2}{\tesla} \left( \frac{a_{\textsc{b}}}{\SI{3.17}{\nano\meter}} \right)^{2} \left( \frac{\omega}{\qty{1}{\tera\hertz}} \right)^{-1} \left( \frac{\mathcal{I}}{\SI{E8}{\watt\per\centi\meter\squared}} \right)\,. 
    \end{multlined}
\end{equation}
Due to the dopants possessing spatially extended states already at low $n$, we begin our calculations at $n=2$ and go up to $n=14$ to illustrate the scaling behavior. Experimental work has studied excitations of these dopants up to $n=10$~\cite{murdin2013,steger2009}, although we are not aware of a particular cutoff of $n$ for these states.
As shown in Fig.~\ref{fig:rydbergife}, the scaling of the IFE for the dopants does not quite scale as $n^4$ like the higher-$n$ states for the Rydberg atoms. For intermediate principal quantum numbers, $n \lesssim10$, the scaling is less obvious.

A magnetization per dopant atom can be obtained from the relation
\begin{equation}
    \mathcal{M}_{\text{eff}} = \frac{\chi}{\mu_0} \mathcal{B}_{\text{eff}} ,
\end{equation}
where $\chi$ is the volume magnetic susceptibility and $\mu_0$ is the permeability of free space.
For silicon, $\chi_{\text{Si}}/\mu_0 \approx \qty{-2.97}{\ampere\squared\second\squared\per\kilogram\per\meter}$~\cite{wolfram}. This leads to an induced magnetization from the IFE as $\mathcal{M}_{\text{eff}} \sim \qty{E3}{\ampere\per\meter}$. For reference we also note the value for germanium, $\chi_{\text{Ge}}/\mu_0 \approx \qty{-6.35}{\ampere\squared\second\squared\per\kilogram\per\meter}$~\cite{wolfram}. As $\mathcal{M} \propto \chi$ and $\chi_{\text{Ge}} > \chi_{\text{Si}}$, we again see that the IFE for germanium is potentially greater than that for silicon.
This estimate is for the magnetization of the atom itself due to orbital motion of electrons under the applied field. The full magnetization of the atom contains paramagnetic contributions from the spins and orbital angular momentum of the bound electrons, as well as a diamagnetic contribution from silicon within the Rydberg atom diameter. The former contributions are present for Rydberg states of both atoms in vacuum and donors in semiconductors, while the latter effect is absent for the free Rydberg atoms. These contributions need to be evaluated separately.

\begin{table}[htp!]
\caption{Diagonal matrix elements of the effective Hamiltonian showing values for the magnitude of the effective magnetic field $\mathcal{B}_{\text{eff}}$ and the corresponding magnetization $\mathcal{M}_{\text{eff}}$ for Si:P for a $\qty{1}{\tera\hertz}$ beam of intensity $\qty{E8}{\watt\per\centi\meter\squared}$. This estimate is for the magnetization of the atom itself and does not take into account both the combined orbital paramagnetism and core diamagnetism.\label{tab:matrixelements}}
\begin{tabular}{|rl|c|c|c|}
\hline
    $|a\rangle =$&$|n,\ell,m\rangle$ &	$\langle a | H_{\text{eff}} | a \rangle / \mu_B$ $[\unit{\tesla}]$ &   $\mathcal{M}_{\text{eff}}$ $[\unit{\ampere\per\meter}]$
\\\hline
    &$| 2,1,\pm1 \rangle$    &	$\mp\num{81}$   &   $\num{\pm240}$
    \\
    &$| 3,1,\pm1 \rangle$    &	$\mp \num{221}$   &   $\num{\pm656}$
    \\
    &$| 3,2,\pm1 \rangle$	&	$\mp\num{146}$   &   $\num{\pm434}$
    \\
    &$| 3,2,\pm2 \rangle$	&	$\mp \num{292}$   &   $\num{\pm867}$
\\\hline
\end{tabular}
\end{table}

The calculation presented here is for the induced magnetization per dopant atom. Here we assume that each dopant can be analyzed independently of each other, so for a bulk material we take as an assumption that the density of the dopants is such that the wave functions of the dopants do not overlap.
A typical concentration of dopants in Si:P is $\qty{E14}{\per\centi\meter\cubed}$. Assuming a uniform distribution of dopants in the volume, this implies an interdopant distance of $d \sim \qty{E-7}{\meter} > \a$. Since this distance is greater than the Bohr radius of the dopants by over an order of magnitude, we expect that samples of this concentration are compatible with our approximations and will allow for the effect we describe.

\section{Conclusion\label{sec:conclusion}}

The inverse Faraday effect has been well studied in the bulk materials. Building upon this understanding,  we focused on IFE applications on shallow semiconductor dopants and Rydberg states.   We demonstrate the possibility of inducing large orbital magnetization IFE in these examples of Rydberg systems.

The first example was that of an isolated atom in a large $n$ Rydberg state, specifically rubidium and cesium atoms in states up to $n=32$, where we found that the IFE can induce an effective magnetic field on the order of $\qty{10}{\milli\tesla}$ for this value of $n$ and a terahertz beam fluence of $\SI{10}{\watt\per\centi\meter\squared}$, and that the effect scales as $n^4$, implying a very large enhancement for higher $n$ states.

The second example studied was that of shallow semiconductor dopants. We found that for phosphorus-doped silicon Si:P using an incident beam fluence of $\SI{E8}{\watt\per\centi\meter\squared}$, the IFE can produce an induced effective magnetic field for the dopant states of $\mathcal{B}_{\text{eff}} \sim O(\qty{E2}{\tesla})$ or a corresponding orbital magnetization of $\mathcal{M}_{\text{eff}} \sim O(\qty{E3}{\ampere\per\meter})$. This process was shown schematically in Fig.~\ref{fig:schematic}.
The values computed for the IFE in this paper have been for silicon as our work is partially motivated by experimental results for Si:P~\cite{vanloon}. Terahertz pumping of silicon is already well established experimentally. However, we wish to remark that germanium would also be of experimental interest as the comparatively higher susceptibility of germanium would further increase the magnitude of the effect. We propose that doped germanium would also be of great experimental interest.

In our examples we considered a driving beam in the terahertz regime. We know that the IFE is a nonresonant effect that scales with frequency, yet terahertz frequency range is not required. To maximize the effect it pays to drive the system close to the transition frequency but it is not necessary. Previous experiments on IFE showed that the effect is maximized near resonance, but still finite over a range of frequencies, such as in the experiments of Ref.~\cite{kozhaev2018}.

We note that the calculation we presented for the above cases is based on the assumption of laser time scales on the order of nanoseconds with effectively stationary magnetization dynamics. In the ultrafast regime of laser pulses on subpicosecond time scales with magnetization dynamics which cannot be assumed to be stationary, a modified analysis is required~\cite{popova}. Indeed, it has been shown that the thermodynamic approach to the IFE cannot describe the effect in the ultrafast regime~\cite{reid}.

In a broader context, Rydberg states have been proposed as platforms for atom-based quantum computing, owing to their long range entanglement properties~\cite{rydbergqi}. In a related development, it has been observed that shallow dopants in semiconductors exhibit characteristics similar to atoms in optical traps. This resemblance opens possibilities for leveraging dopants in semiconductors for quantum computing~\cite{greenland}.
It has previously been demonstrated that ultrafast switching of magnetic moments is possible via the IFE~\cite{vahaplar,lottermoser,kimel}.
Inducing the IFE in doped silicon, as we propose in this paper, could therefore have ramifications for dopant-based quantum computing~\cite{fricke}, with the IFE having the potential to coherently control the qubit states of dopant qubits in semiconductors. Specifically, the IFE has the potential to prepare cat states~\cite{raimond2001} by polarizing the dopant qubits.

\begin{acknowledgments}
We are grateful to R.~Castillo-Garza, M.~Jain,  C.~Ahn, I.~Sochnikov, D.~Vu, and F.~Walker for discussions on possible experimental realizations of these ideas, as well as the referees for their useful comments. We wish to thank T.-T.~Yeh for assistance with illustrations. This material is based upon work supported by the U.S. Department of Energy, Office of Science, Office of Basic Energy Sciences under Award Number DE-SC-0025580 (AVB and PJW -- conceptualization, calculation, writing), the European Research Council under the European Union Seventh Framework Project No. ERS-2018-SYG 810451 HERO (AVB and GA -- conceptualization, writing), and the University of Connecticut (PJW -- travel).
\end{acknowledgments}

\appendix

\begin{widetext}

\section{Derivation of the effective Hamiltonian\label{sec:derivation}}

In this appendix we review the derivation of the effective Hamiltonian employed in the main text to describe the inverse Faraday effect. Our derivation follows the construction originally developed in Ref.~\cite{pershan}.
The wave function for the dopants obeys the evolution equation
\begin{equation}
    -\frac{\hbar}{i} \frac{\partial \Psi(t)}{\partial t} = H \Psi(t) ,
\end{equation}
where we consider a Hamiltonian of the form $H = H_0 + V(t)$, with $H_0$ the unperturbed Hamiltonian and $V(t)$ an interaction with an external field to be considered perturbatively. Spin-orbit coupling is not considered in the following.
In the interaction representation, the evolution equation takes the form
\begin{equation}
    -\frac{\hbar}{i} \frac{\partial \overline{\Psi}(t)}{\partial t} = \overline{V}(t) \overline{\Psi}(t)
\end{equation}
where $\overline{\Psi}(t) = e^{i H_0 t /\hbar} \Psi(t)$ and $\overline{V}(t) = e^{i H_0 t/\hbar} V(t) e^{-i H_0 t/\hbar}$. 
The formal solution to the evolution equation can be obtained as
\begin{align}
	\Psi(t)
	&=
    e^{-i H_0 t/\hbar}
	\left[ 1 - \frac{i}{\hbar} \int_{-\infty}^{t} dt' \overline{V}(t') - \frac{1}{\hbar^2} \int_{-\infty}^{t} dt' \overline{V}(t') \int_{-\infty}^{t'} dt'' \overline{V}(t'') + O(\hbar^{-3}) \right] \Psi({0}) \,.
\end{align}
We define an effective Hamiltonian $\overline{H}_{\text{eff}}(t)$ in the interaction representation as the generator of transitions between states $|a\rangle$ and $|b\rangle$ due to the interaction $V(t)$ by
\begin{equation}
    \left\langle a \left| 1 - \frac{i}{\hbar} \int_{-\infty}^{t} dt' \overline{H}_{\text{eff}}(t') \right| b \right\rangle
    =
    \left\langle a \left| 1 - \frac{i}{\hbar} \int_{-\infty}^{t} dt' \overline{V}(t') - \frac{1}{\hbar^2} \int_{-\infty}^{t} dt' \overline{V}(t') \int_{-\infty}^{t'} dt'' \overline{V}(t'') + O(\hbar^{-3}) \right| b \right\rangle \,.
\label{eq:perturbationexpansion}
\end{equation}
The term of interest here is the process second-order in $V(t)$. The first-order term is neglected as we assume that there are no relevant absorption processes, assuming the optical frequency is non-resonant with any atomic frequency, so we have $\langle a | \overline{V}(t') | b \rangle = 0$.
This leads to the effective Hamiltonian being defined in terms of the matrix elements of the second-order interaction as
\begin{align}
	\langle a | \overline{H}_{\text{eff}}(t) | b \rangle &\vcentcolon= \frac{i}{\hbar} \sum_{c} \langle a | \overline{V}(t) | c \rangle \int_{-\infty}^{t} \langle c | \overline{V}(t') | b \rangle d t' \,.
\label{eq:effhamV}
\end{align}
Here the states $|a\rangle$ and $|b\rangle$ are eigenstates of the ground multiplet and $|c\rangle$ are excited intermediate states.
The integral can be evaluated by using the approximation that the interaction is slowly varying over the timescale considered. Assuming variations in the interaction $v(t)$ occur on a characteristic time scale of $\tau$, the integral in Eq.~\eqref{eq:effhamV} can be approximated with
\begin{align}
    \int_{-\infty}^{t} \langle c | \overline{V}(t') | b \rangle d t'
    &\approx
    \langle c | v(t) | b \rangle \frac{e^{i (\omega_{cb} + \omega) t}}{i (\omega_{cb} + \omega)}
    +
    \langle c | v^*(t) | b \rangle \frac{e^{i (\omega_{cb} - \omega) t}}{i (\omega_{cb} - \omega)} \,.
\end{align}
This approximation is valid in the case where the characteristic time scale is in the regime where $\tau |\omega \pm \omega_{ab}| \gg 1$, such that $v(t)$ can be considered constant over the integration time, with an adiabatic ramp-up initial condition $v(-\infty) = 0$. With the understanding of this approximation, we suppress the time dependence of $v$ and $v^*$ in the following.
This approximation also assumes that the transitions are off-resonant, $|\omega_{ab}\pm\omega| \;\slashed\ll\; \omega_{ab}$. Here we define the notation $\hbar \omega_{ab} \equiv E_a - E_b$ where $E_a$ is the energy of state $a$.
Using this approximation we can write the effective Hamiltonian as
\begin{align}
    \langle a | \overline{H}_{\text{eff}}(t) | b \rangle
    &=  \frac{i}{\hbar} \sum_{c} \langle a | \overline{V}(t) | c \rangle \int_{-\infty}^{t} \langle c | \overline{V}(t') | b \rangle d t'
    \\
    &\approx \frac{i}{\hbar} \sum_{c} \left( e^{i (\omega_{ac} + \omega) t} \langle a | v | c \rangle + e^{i (\omega_{ac} - \omega) t} \langle a | v^* | c \rangle \right)
        \left( \frac{e^{i (\omega_{cb} + \omega) t}}{i (\omega_{cb} + \omega)} \langle c | v | b \rangle + \frac{e^{i (\omega_{cb} - \omega) t}}{i(\omega_{cb} - \omega)} \langle c | v^* | b \rangle \right)
    \\
    &=  \begin{multlined}[t]
    \frac{1}{\hbar} \sum_{c} \left[ \frac{\langle a |v| c \rangle \langle c |v^*| b \rangle}{\omega_{bc} - \omega} + \frac{\langle a |v^*| c \rangle \langle c |v| b \rangle}{\omega_{bc} + \omega} 
    +
    \frac{\langle a |v| c \rangle \langle c |v| b \rangle}{\omega_{bc} + \omega} e^{i 2\omega t}
    +
    \frac{\langle a |v^*| c \rangle \langle c |v^*| b \rangle}{\omega_{bc} - \omega} e^{-i 2\omega t} \right] e^{i \omega_{ab} t} \label{eq:zhg}
    \end{multlined}
\end{align}
The second pair of terms on the right-hand side of Eq.~\eqref{eq:zhg} are proportional to $e^{\pm i 2\omega t}$ and therefore represent second-order harmonic generation processes. These terms will not be considered in the following as our focus in this paper is only on the zeroth harmonic generation (dc) effects.

The interaction we consider is the electric dipole interaction, with the interaction taking the form of
\begin{align}
    v &= -q \left( r_+ \mathcal{E}_R + r_- \mathcal{E}_L \right) \,,
    &
    v^* &= -q \left( r_- \mathcal{E}_R^* + r_+ \mathcal{E}_L^* \right) \,.
\end{align}
The rectified zeroth-order harmonic mode can now be expanded as
\begin{align}
    \langle a | \overline{H}_{\text{eff}}(t) | b \rangle
	&= \frac{1}{\hbar} \sum_{c} \left[ \frac{\langle a | v | c \rangle \langle c | v^* | b \rangle}{\omega_{bc} + \omega} + \frac{\langle a | v^* | c \rangle \langle c | v | b \rangle}{\omega_{bc} - \omega} \right] e^{i \omega_{ba} t}
    \\
    &=  \frac{1}{\hbar} \sum_{c} \left[ \frac{(\omega_{bc} + \omega)\langle a | v | c \rangle \langle c | v^* | b \rangle}{\omega_{bc}^2 - \omega^2} + \frac{(\omega_{bc} - \omega)\langle a | v^* | c \rangle \langle c | v | b \rangle}{\omega_{bc}^2 - \omega^2} \right] e^{i \omega_{ba} t}
    \\
    &=  \begin{multlined}[t]
    \frac{q^2}{\hbar} \sum_{c} \bigg[ \frac{(\omega_{bc} + \omega)\langle a | r_+ \mathcal{E}_R + r_- \mathcal{E}_L | c \rangle \langle c | r_- \mathcal{E}_R^* + r_+ \mathcal{E}_L^* | b \rangle}{\omega_{bc}^2 - \omega^2}
    \\+
    \frac{(\omega_{bc} - \omega)\langle a | r_- \mathcal{E}_R^* + r_+ \mathcal{E}_L^* | c \rangle \langle c | r_+ \mathcal{E}_R + r_- \mathcal{E}_L | b \rangle}{\omega_{bc}^2 - \omega^2} \bigg] e^{i \omega_{ba} t}
    \end{multlined}
    \\
    &=  \begin{aligned}[t]
    \frac{q^2}{\hbar} \sum_{c} \bigg[
    &(\mathcal{E}_R^2 -  \mathcal{E}_L^2) \omega \frac{\langle a | r_+ | c \rangle \langle c | r_- | b \rangle -
    \langle a | r_- | c \rangle \langle c | r_+ | b \rangle}{\omega_{bc}^2 - \omega^2}
    \\&+
    (\mathcal{E}_R^2 + \mathcal{E}_L^2) \omega_{bc}
    \frac{\langle a | r_+ | c \rangle \langle c | r_- | b \rangle +
    \langle a | r_- | c \rangle \langle c | r_+ | b \rangle}{\omega_{bc}^2 - \omega^2}
    \\&+
    2 \mathcal{E}_R \mathcal{E}_L^* \omega_{bc}
    \frac{\langle a | r_+ | c \rangle \langle c | r_+ | b \rangle}{\omega_{bc}^2 - \omega^2}
    +
    2 \mathcal{E}_L \mathcal{E}_R^* \omega_{bc}
    \frac{\langle a | r_- | c \rangle \langle c | r_- | b \rangle}{\omega_{bc}^2 - \omega^2}
    \bigg] e^{i \omega_{ab} t} \,.
    \end{aligned}
\end{align}
Expansion of the matrix elements of $v$ and $v^*$ yields contributions to the effective Hamiltonian which are proportional to $\omega (\mathcal{E}_R^2 - \mathcal{E}_L^2)$, $\omega_{bc} (\mathcal{E}_R^2 + \mathcal{E}_L^2)$, and $\omega_{bc} (\mathcal{E}_R \mathcal{E}_L^* + \mathcal{E}_L \mathcal{E}_R^*)$. In addition, $\mathcal{E}_{L/R}^2 = \mathcal{E}_{L/R}^* \mathcal{E}_{L/R}$. For transitions involving $\Delta n = 0$, the energies $\omega_{b}$ and $\omega_{c}$ are degenerate, so these second two terms do not contribute. For transitions where $\Delta n \neq 0$, $\omega_{bc} \sim 10^{-2} \omega$ (for $\omega \sim \qty{1}{\tera\hertz}$), so these terms again do not significantly contribute compared to the leading-order $\mathcal{E}_R^2 - \mathcal{E}_L^2$ term.

The final expression we utilize for the effective Hamiltonian matrix elements used in the main text [Eq.~\eqref{eq:effham}] is then
\begin{align}
    \langle a | H_{\text{eff}} | b \rangle
	&= \frac{q^2}{\hbar}
	\sum_c \left( \frac{\langle a |r_+| c \rangle \langle c |r_-| b \rangle }{\omega_{bc}^2 - \omega^2} - \frac{\langle a |r_-| c \rangle \langle c |r_+| b \rangle }{\omega_{bc}^2 - \omega^2} \right) \omega \left( \mathcal{E}_R^2 - \mathcal{E}_L^2 \right) \,.
\label{eq:effhamapp}
\end{align}
Here $H_{\text{eff}}$ is obtained from $\overline{H}_{\text{eff}}(t) = e^{i H_0 t/\hbar} H_{\text{eff}} e^{-i H_0 t/\hbar}$ such that $\langle a | \overline{H}_{\text{eff}}(t) | b \rangle = \langle a | {H}_{\text{eff}} | b \rangle e^{i \omega_{ab} t}$.
As defined in Eq.~\eqref{eq:perturbationexpansion}, the full dynamical expression for the transition amplitude is given by $\left\langle a \left| \int_{-\infty}^{t} d t' \overline{H}_{\text{eff}}(t') \right| b \right\rangle$. The integration yields
\begin{equation}
    \left\langle a \left| \int_{-\infty}^{t} d t' \overline{H}_{\text{eff}}(t') \right| b \right\rangle
    =
    \frac{1}{i \omega_{ab}} \langle a | {H}_{\text{eff}} | b \rangle e^{i \omega_{ab} t} \,.
\end{equation}
We may appropriately use the expression \eqref{eq:effhamapp} by assuming a timescale such that $\langle a | {H}_{\text{eff}} | b \rangle e^{i \omega_{ab} t} / (i \omega_{ab}) \approx \tau \langle a | {H}_{\text{eff}} | b \rangle$ where $\tau$ is a time scale.

The set of states $|a\rangle$, $|b\rangle$, and $|c\rangle$ in Eq.~\eqref{eq:effhamapp} consists of eigenstates of the unperturbed Hamiltonian $H_0$.
The Rydberg behavior of the excited atomic systems and the semiconductor dopant states means that they can be approximated with
hydrogenic wave functions with atomic number $Z$ and Bohr radius $\a$:
\begin{equation}
    |a\rangle = | n,\ell,m\rangle = \sqrt{\left( \frac{2 Z}{n a_{\textsc{b}}} \right)^3 \frac{(n-\ell-1)!}{2 n (n+\ell)!}} e^{-\frac{Z r}{n a_{\textsc{b}}}} \left( \frac{2 Z r}{n a_{\textsc{b}}} \right)^\ell L^{2\ell+1}_{n-\ell-1}(\tfrac{2 Z r}{n a_{\textsc{b}}}) Y_{\ell,m}(\theta,\phi) \,.
\label{eq:hydrogenicwf}
\end{equation}
The energies of the eigenstates are given by
\begin{equation}
    E_{n,\ell,m} = -\frac{Z^2 q^2}{8 \pi \epsilon a_{\textsc{b}}} \frac{1}{n^2} \left[ 1 + \frac{\alpha^2}{n^2} \left( \frac{n}{\ell + \frac12} - \frac34 \right) \right]
\end{equation}
where $\epsilon$ is the dielectric constant and $\alpha = q^2/4\pi \epsilon \hbar c$ is the fine-structure constant.
The matrix elements are calculated considering wave functions of the form \eqref{eq:hydrogenicwf} in spherical coordinates with
$r_\pm = \frac{1}{\sqrt{2}} ( x \pm i y) = \frac{1}{\sqrt{2}} r \sin\theta (\cos\phi \pm i \sin\phi)$. 

\end{widetext}

\section{Numerical data for Rydberg atoms\label{sec:rydbergdata}}

Presented in this appendix are the numerical values of the induced effective magnetic field for Rydberg states plotted in Fig.~\ref{fig:rydbergife} for rubidium (Table~\ref{tab:ruife}) and cesium (Table~\ref{tab:csife}). We calculate the effective magnetic field from the matrix element $\langle a|H_{\text{eff}}| a\rangle$ with $|a\rangle = |n,1,\pm1\rangle$. The intermediary states we sum over are $|c\rangle \in \{ |n,0,0\rangle , |n,2,\pm2\rangle \}$.
The fitted functions for the effective magnetic field in Fig.~\ref{fig:rydbergife} are 
\begin{multline}
\mathcal{B}_{\text{Rb}}(n) = -\num{4.21189E-5} + \num{1.6737E-5} n -\\- \num{2.3217E-6} n^2 + \num{9.33475E-8} n^3 +\\+ \num{1.44179E-8} n^4
\label{eq:rbfit}
\end{multline}
and
\begin{multline}
\mathcal{B}_{\text{Cs}}(n) = -\num{7.08553E-5} + \num{1.90833E-5} n -\\- \num{1.97067E-6} n^2 + \num{7.15958E-8} n^3 +\\+ \num{7.77412E-9} n^4 ,
\label{eq:csfit}
\end{multline}
which have $n^4$ at dominant order for large $n$.
\begin{table}[htp!]
\caption{Magnitude of the induced effective magnetic field by the IFE for the $|n,1,\pm1\rangle$ state of Rb induced by a $\qty{1}{\tera\hertz}$ beam of intensity $\qty{10}{\watt\per\centi\meter\squared}$.\label{tab:ruife}}
\begin{tabular}{| r | c |}
\hline
    $n$ &   $\mathcal{B}_\text{eff}$ $[\unit{\milli\tesla}]$ \\
    \hline
    $8$ &   $\pm \num{4.58E-2}$ \\\hline
    $10$ &   $\pm \num{1.28E-1}$ \\\hline
    $12$ &   $\pm \num{2.86E-1}$ \\\hline
    $14$ &   $\pm \num{5.50E-1}$ \\\hline
    $16$ &   $\pm \num{9.62E-1}$ \\\hline
    $18$ &   $\pm \num{1.57}$ \\\hline
    $20$ &   $\pm \num{2.42}$ \\\hline
    $22$ &   $\pm \num{3.57}$ \\\hline
    $24$ &   $\pm \num{5.09}$ \\\hline
    $26$ &   $\pm \num{7.05}$ \\\hline
    $28$ &   $\pm \num{9.51}$ \\\hline
    $30$ &   $\pm \num{12.6}$ \\\hline
    $32$ &   $\pm \num{16.3}$ \\
\hline
\end{tabular}
\end{table}
\begin{table}[htp!]
\caption{Magnitude of the induced effective magnetic field by the IFE for the $|n,1,\pm1\rangle$ state of Cs induced by a $\qty{1}{\tera\hertz}$ beam of intensity $\qty{10}{\watt\per\centi\meter\squared}$.\label{tab:csife}}
\begin{tabular}{| r | c |}
\hline
    $n$ &   $\mathcal{B}_\text{eff}$ $[\unit{\milli\tesla}]$ \\
    \hline
    $8$ &   $\pm \num{2.52E-2}$ \\\hline
    $10$ &   $\pm \num{7.12E-2}$ \\\hline
    $12$ &   $\pm \num{1.58E-1}$ \\\hline
    $14$ &   $\pm \num{3.05E-1}$ \\\hline
    $16$ &   $\pm \num{5.33E-1}$ \\\hline
    $18$ &   $\pm \num{8.69E-1}$ \\\hline
    $20$ &   $\pm \num{1.34}$ \\\hline
    $22$ &   $\pm \num{1.97}$ \\\hline
    $24$ &   $\pm \num{2.82}$ \\\hline
    $26$ &   $\pm \num{3.90}$ \\\hline
    $28$ &   $\pm \num{5.28}$ \\\hline
    $30$ &   $\pm \num{6.96}$ \\\hline
    $32$ &   $\pm \num{9.02}$ \\
\hline
\end{tabular}
\end{table}

\section{Numerical data for shallow dopants\label{sec:dopantdata}}

In this appendix we give the numerical data of the IFE calculations for the shallow silicon dopants presented in Sec.~\ref{sec:ifeshallowdopants} and plotted in Fig.~\ref{fig:rydbergife}.
The intermediate states which are summed over are those which obey the selection rules $\langle n',\ell\pm1,m\pm1 |r_\pm| n,\ell,m \rangle$, and we take $n' \in \{n,n\pm1\}$. Terms not obeying these rules vanish, either because their overlap is zero, or because summing over intermediary states with $\pm m$ results in values which are equal in magnitude and opposite in sign, and therefore cancel.
Transitions with $\Delta n > 1$ are finite, but of subleading order and so are not taken into account in our calculations.
The diagonal matrix elements of the effective Hamiltonian including all intermediate virtual states considered are shown in Table~\ref{tab:allmatrixelements}.

As with the effective magnetic field generated by the Rydberg atoms, the shallow dopants also exhibit scaling with $n^4$.
The IFE for the dopants was computed for states of the form $|n,1,\pm1\rangle$, $|n,2,\pm1\rangle$, and $|n,2,\pm2\rangle$. Curves fitted to the numerical data for these states as plotted in Fig.~\ref{fig:rydbergife} are
\begin{multline}
    \mathcal{B}_{|n,1,\pm1\rangle}(n)
    =
    148.914 - 148.675 n + 67.1985 n^2 -\\- 4.06475 n^3 + 0.144196 n^4 ,
\label{eq:sip11}
\end{multline}
\begin{multline}
    \mathcal{B}_{|n,2,\pm1\rangle}(n)
    =
    142.125 - 163.611 n + 64.4927 n^2 -\\- 3.31341 n^3 + 0.105551 n^4 ,
\label{eq:sip21}
\end{multline}
and
\begin{multline}
    \mathcal{B}_{|n,2,\pm2\rangle}(n)
    =
    385.522 - 388.863 n + 141.487 n^2 -\\- 7.67950 n^3 + 0.242497 n^4 .
\label{eq:sip22}
\end{multline}

\begin{widetext}

\begin{table}[htp!]
\caption{Effective magnetic field values of Si:P for the diagonal matrix elements $| n,1,\pm1 \rangle$, $| n,2,\pm1 \rangle$, and $| n,2,\pm2 \rangle$ induced by a $\qty{1}{\tera\hertz}$ beam of intensity $\qty{E8}{\watt\per\centi\meter\squared}$. \label{tab:allmatrixelements}}
\begin{tabular}{|c|c|c|c|}
\hline
    $n$	&   $\langle n,1,\pm1 | H_{\text{eff}} | n,1,\pm1 \rangle / \mu_B$ $[\unit{\tesla}]$	&	$\langle n,2,\pm1 | H_{\text{eff}} | n,2,\pm1 \rangle / \mu_B$ $[\unit{\tesla}]$	&	$\langle n,2,\pm2 | H_{\text{eff}} | n,2,\pm2 \rangle / \mu_B$ $[\unit{\tesla}]$
\\\hline
2	&	$\mp81$	&	---	&	---
\\\hline
3	&	$\mp221$	&	$\mp146$	&	$\mp292$
\\\hline
4	&	$\mp415$	&	$\mp342$	&	$\mp684$
\\\hline
5	&	$\mp671$	&	$\mp593$	&	$\mp1180$
\\\hline
6	&	$\mp967$	&	$\mp899$	&	$\mp1800$
\\\hline
7	&	$\mp1350$	&	$\mp1270$	&	$\mp2520$
\\\hline
8	&	$\mp1750$	&	$\mp1680$	&	$\mp3360$
\\\hline
9	&	$\mp2260$	&	$\mp2190$	&	$\mp4380$
\\\hline
10	&	$\mp2770$	&	$\mp2700$	&	$\mp5400$
\\\hline
11	&	$\mp3350$	&	$\mp3280$	&	$\mp6560$
\\\hline
12	&	$\mp4010$	&	$\mp3930$	&	$\mp7850$
\\\hline
13	&	$\mp4730$	&	$\mp4640$	&	$\mp9270$
\\\hline
14	&	$\mp5640$	&	$\mp5460$	&	$\mp\num{10930}$
\\\hline
\end{tabular}
\end{table}

\end{widetext}

\bibliography{ife}

\end{document}